\documentclass[twocolumn]{aastex7}

\usepackage{float}
\usepackage{stfloats}

\usepackage{hyperref}
\usepackage{amsmath}
\usepackage{graphicx, caption}
\usepackage{siunitx}
\usepackage{xcolor}
\usepackage{fancyhdr}
\usepackage{booktabs}
\usepackage{natbib}
\usepackage[export]{adjustbox}
\usepackage{nidanfloat}  
\usepackage{stfloats}
\usepackage{wrapfig}
\usepackage[flushleft]{threeparttable}
\usepackage{changepage}
\usepackage{subcaption}
\usepackage{caption}
\usepackage{courier}
\renewcommand\ttfamily{\fontfamily{pcr}\selectfont}
\newcommand{\Msol}{\hbox{M$_\sun$}}

\setcitestyle{authoryear,open={(},close={)}}

\newcommand{\del}[1]{\relax}%
\newcommand{\DELETED}[1]{\relax}%
{\relax}%
\newcommand\eg           {{\it e.g.}, }

\begin{document}

\title{Constraints on the Pop III Sky Surface Brightness from High-Redshift Caustic Transits in MACS0416}

\pagestyle{fancy}
\fancyhf{}
\renewcommand{\headrulewidth}{0pt}
\fancyfoot[C]{\thepage}


\author[0009-0000-3166-0595]{Jakob P. Perivolotis}
\email{jpcg4@missouri.edu}
\affiliation{Department of Physics and Astronomy, University of Missouri,
Columbia, MO 65211, USA}
\affiliation{School of Earth and Space Exploration, Arizona State University,
Tempe, AZ 85287-1404, USA}

\author[0000-0003-1383-9414]{Pietro Bergamini}
\email{pietro.bergamini@unimi.it}
\affiliation{Dipartimento di Fisica, Università degli Studi di Milano, via Celoria 16, I-20133 Milano, Italy}

\author[0000-0001-8156-6281]{Rogier A.\ Windhorst} \email{Rogier.Windhorst@gmail.com}
\affiliation{School of Earth and Space Exploration, Arizona State University,
Tempe, AZ 85287-1404, USA}

\author[0000-0003-1096-2636]{Erik Zackrisson}
\email{erik.zackrisson@physics.uu.se}
\affiliation{Department of Physics and Astronomy, Uppsala University, Box 516, SE-751 20 Uppsala, Sweden}

\author[0000-0001-6650-2853]{Timothy Carleton}  \email{tmcarlet@asu.edu}
\affiliation{School of Earth and Space Exploration, Arizona State University,
Tempe, AZ 85287-1404, USA}

\author[0000-0003-3329-1337]{Seth H.\ Cohen}  \email{seth.cohen@asu.edu}
\affiliation{School of Earth and Space Exploration, Arizona State University,
Tempe, AZ 85287-1404, USA}

\author[0000-0002-9816-1931]{Jordan C.\ J.\ D'Silva}  \email{jordan.dsilva@research.uwa.edu.au}
\affiliation{International Centre for Radio Astronomy Research (ICRAR) and the
International Space Centre (ISC), The University of Western Australia, M468,
35 Stirling Highway, Crawley, WA 6009, Australia}
\affiliation{ARC Centre of Excellence for All Sky Astrophysics in 3 Dimensions
(ASTRO 3D), Australia}

\author[0000-0002-9984-4937]{Rachel Honor} \email{rchonor@asu.edu}
\affiliation{School of Earth and Space Exploration, Arizona State University, Tempe, AZ 85287-1404, USA}

\author[0000-0003-1268-5230]{Rolf A.\ Jansen} \email{rolfjansen.work@gmail.com}
\affiliation{School of Earth and Space Exploration, Arizona State University,
Tempe, AZ 85287-1404, USA}

\author[0000-0002-6610-2048]{Anton M.\ Koekemoer} \email{koekemoer@stsci.edu}
\affiliation{Space Telescope Science Institute,
3700 San Martin Drive, Baltimore, MD 21218, USA}

\author[0000-0002-6150-833X]{Rafael {Ortiz~III}} \email{rortizii@asu.edu}
\affiliation{School of Earth and Space Exploration, Arizona State University,
Tempe, AZ 85287-1404, USA}

\author[0000-0002-7265-7920]{Jake Summers} \email{jssumme1@asu.edu}
\affiliation{School of Earth and Space Exploration, Arizona State University,
Tempe, AZ 85287-1404, USA}

\author[0000-0001-7410-7669]{Dan Coe} \email{dcoe@stsci.edu}
\affiliation{Space Telescope Science Institute, 3700 San Martin Drive, Baltimore, MD 21218, USA}
\affiliation{Association of Universities for Research in Astronomy (AURA) for the European Space Agency (ESA), STScI, Baltimore, MD 21218, USA}
\affiliation{Center for Astrophysical Sciences, Department of Physics and Astronomy, The Johns Hopkins University, 3400 N Charles St. Baltimore, MD 21218, USA}

\author[0000-0003-1949-7638]{Christopher J.\ Conselice} \email{conselice@gmail.com}
\affiliation{Jodrell Bank Centre for Astrophysics, Alan Turing Building,
University of Manchester, Oxford Road, Manchester M13 9PL, UK}

\author[0000-0001-9065-3926]{Jose M. Diego} \email{chemadiegor@gmail.com}
\affiliation{Instituto de F\'isica de Cantabria (CSIC-UC). Avenida. Los Castros
s/n. 39005 Santander, Spain}


\author[0000-0003-1625-8009]{Brenda Frye} \email{brendafrye@gmail.com}
\affiliation{Department of Astronomy/Steward Observatory, University of Arizona, 933 N Cherry Ave,
Tucson, AZ, 85721-0009, USA}

\author[0000-0003-4223-7324]{Massimo Ricotti}
\email{ricotti@umd.edu}
\affiliation{Department of Astronomy, University of Maryland, College Park, MD 20742, USA}

\author[orcid=0009-0007-0782-0721]{Gibson B.\ Bowling}
\email{gbbowlin@asu.edu}
\affiliation{School of Earth and Space Exploration, Arizona State University, Tempe, AZ 85287-1404, USA}






\author[0000-0002-0648-1699]{Brent Smith}
\affiliation{School of Earth \& Space Exploration, Arizona State University, Tempe, AZ 85287-1404, USA}\email{bsmith18@asu.edu}

\author[0000-0001-9262-9997]{Christopher N.\ A.\ Willmer} \email{cnawillmer@gmail.com}
\affiliation{Steward Observatory, University of Arizona,
933 N Cherry Ave, Tucson, AZ, 85721-0009, USA}


\author[0000-0001-7592-7714]{Haojing Yan} \email{yanhaojing@gmail.com}
\affiliation{Department of Physics and Astronomy, University of Missouri,
Columbia, MO 65211, USA}

\begin{abstract}    

\indent Population III (Pop III) stars are the first hypothetical stellar structures in the universe. These zero metallicity stars, formed from pristine hydrogen and helium left over from the Big Bang nucleosynthesis, are theorized to span a wide range of masses, extending up to several $100\,M_{\odot}$. These massive stars had extremely short lives and are believed to have played a crucial role in the universe's nucleosynthesis and reionization. Their expected fluxes are far below the typical JWST NIRCam detection limits making direct observation highly unlikely. However, by taking advantage of massive foreground galaxy clusters and their caustics, where magnification peaks dramatically, we attempt to detect possible overlooked high-redshift (\(7\leq z\leq17\)) caustic transits in the lensing cluster MACS J0416.1$-$2403. By analyzing three observations spanning 126 days, we create difference-images to identify potential caustic transit candidates. The critical curves, for background sources between $7\leq z \leq17$, derived from a strong lensing model, guided our visual inspection for transits in that redshift range. After examining three difference-image combinations, no additional transits were found. While the longer caustics in our model sweep a larger source-plane area, thereby increasing the probability of detecting an event, we still observe zero transits. Factoring in this increased statistical sensitivity alongside our deeper imaging, we establish a fainter limit for the underlying unresolved stellar population at z$\gtrsim$7. Using this null result, we constrain the 2 \(\mu\)m Pop III sky Surface Brightness (SB) to \(\geq32.8\pm0.6\) mag \(\text{arcsec}^{-2}\). Modeling the non-detection as a Poisson process gives a posterior mean caustic transit rate of 0.29 cluster\(^{-1}\) year\(^{-1}\) and a 95\% upper credible limit of \(\lambda_{95}=0.86\ \mathrm{cluster}^{-1}\,\mathrm{yr}^{-1}\). The inferred rate remains consistent with the adopted fiducial Pop III caustic-transit model and provides an empirical benchmark for future multi-epoch monitoring campaigns. 

\end{abstract}

\section{Introduction}

\indent The James Webb Space Telescope (JWST) marks a new era in astronomical observation, offering unprecedented sensitivity and precision in the infrared spectrum. Its remarkable capabilities extend our understanding of the universe's earliest epochs. Its observations have already challenged existing models of early galaxy formation. The Cosmic Evolution Early Release Science (CEERS) program (PI: S. L. Finkelstein; PID: 1345) has shown that the stellar mass density in galaxies out to $z$ \(\approx\) 6 significantly exceeds theoretical predictions \citep[e.g.][]{Labbe2023, Eyles2007}. These observations suggest star-formation began earlier and proceeded more efficiently than previously thought. Some of these galaxies reached stellar masses comparable to our Milky Way in just 700 million years, whereas the Milky Way began forming only 800 million years after the Big Bang \citep{Xiang2022}. This significant discrepancy between observation and theory necessitates a fundamental revision of our understanding of early galaxy formation and evolution.

\begin{figure*}[t]
\centering
\includegraphics[width=0.90\textwidth]{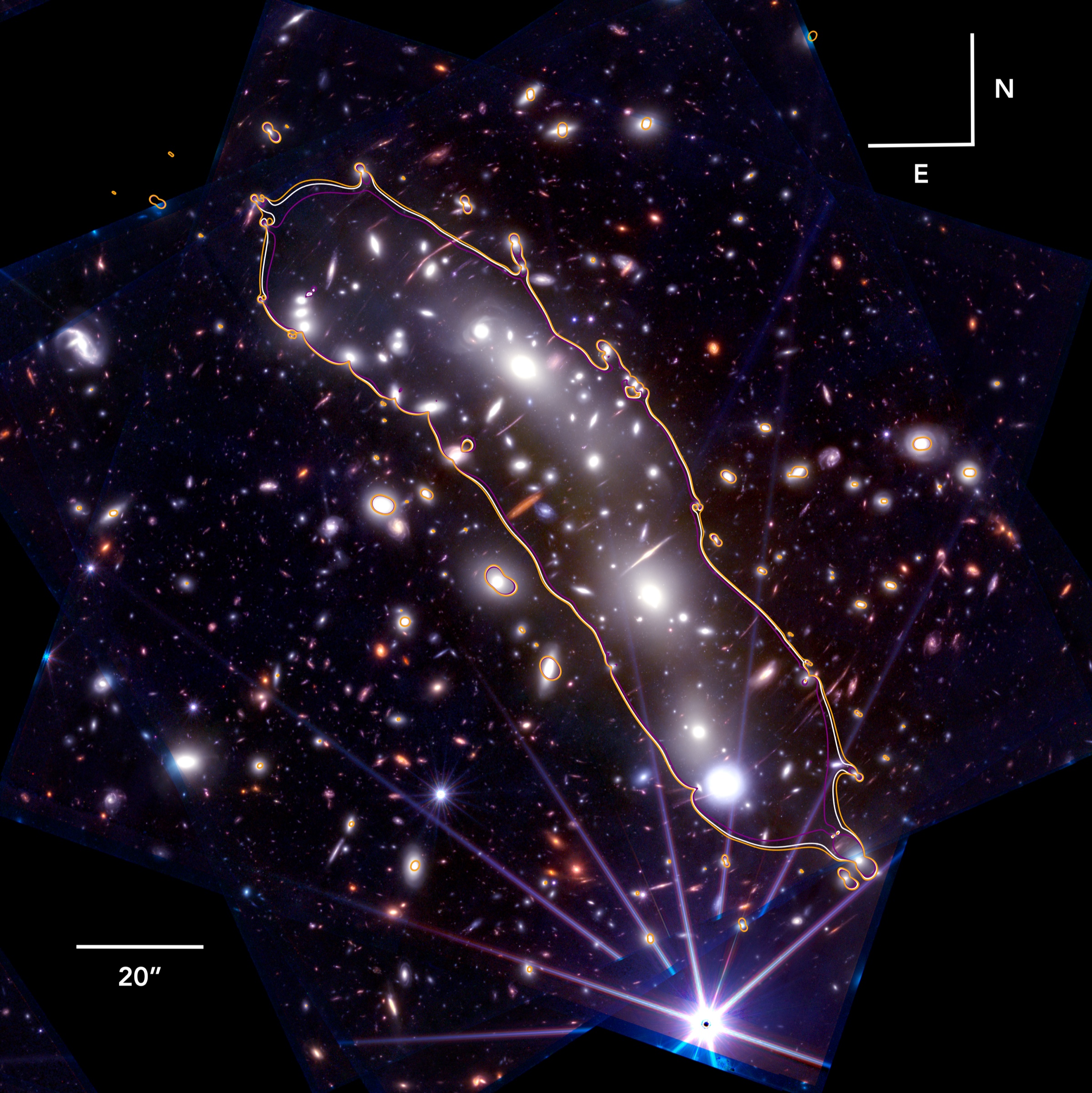}
\caption{False-color composite of MACS0416 using all three PEARLS NIRCam observations at 30mas. The image is color-coded as Red: F444W+F410M+F356W, Green: F277W+F200W, and Blue: F150W+F115W+F090W. Critical curves for sources at redshifts 7, 12, and 17 are shown in purple, white, and orange, respectively.}
\label{fig:color}
\end{figure*}

\vfill\break The incredible depth that JWST can reach is much increased when pointed at a massive lensing galaxy cluster.  These clusters act as a magnifying glass, providing a typical global magnification of \(\mu=10\) \citep{Lotz2017}. The magnification fluctuates due to things like substructures, mass distribution, and proximity to caustics. With \(\mu = 10\), an object's apparent magnitude becomes brighter by 2.5 magnitudes from the relation, \(\Delta m = -2.5\log_{10}(\mu)\), which is significant \citep{Fudamoto2025, Palencia2026}, but not enough to bring structures from the earliest epochs including Pop III stars into view. By taking advantage of the caustics of these natural lenses, where the largest magnification can occur for the smallest point sources, we can obtain a very significant increase in the magnification. 

\begin{figure*}[t]
\centering
\includegraphics[scale=0.75]{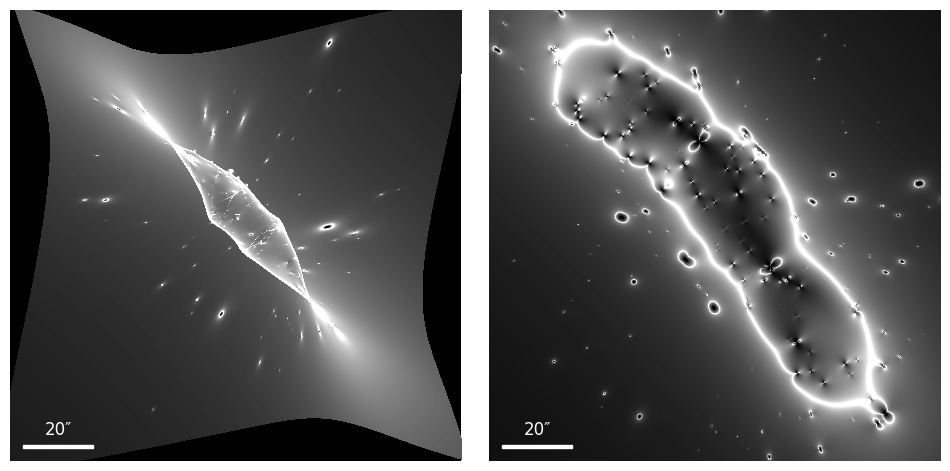}
\caption{Left panel: Source plane caustic map for a background source at $z=10$, generated from a MACS0416 lens model \citep{Bergamini2023a}. Sources crossing these caustics can undergo extreme magnification.
Right panel: Corresponding image plane magnification map for MACS0416 at $z=10$. Bright white regions delineate critical curves, areas of extreme magnification where $\mu$ can be $\geq500$, and darker regions are areas of low magnification \(\mu\approx1\). Both panels are on the same angular scale, but represent the source and image plane, respectively.}
\label{fig:lensing_panel}
\end{figure*}

However, such extreme image magnification is difficult to observe in nature as substructures and microlenses within the lens plane can positively or negatively affect \(\mu\). While a more conservative magnification coeffecient of $\mu = 100 - 1000$ provides a substantial brightness boost of $5-7.5$ mag, it remains insufficient to detect individual stars from the Epoch of First Light. Based on synthetic models of Pop III stars, \eg\ a $100\,M_{\odot}$ star at $z=12$, has an unlensed apparent magnitude of 38.22 mag \citep[\eg][]{Windhorst2018}. To bridge this gap in brightness and bring the star within JWST NIRCam depths of 28.5 mag, a magnitude boost of 9.7 mag would be required, demanding an extreme magnification of $\mu \gtrsim 7.7\times 10^3$. Fortunately, the extreme magnification localized precisely on cluster caustics can range between \(\mu = 10^4 - 10^5\) \citep{Miralda-Escude1991, Kelly2017, Kelly2018} if no fore-ground microlensing is present to dilute this signal \citep{Diego2018}.

The Hubble Space Telescope (HST) remains a valuable complement to JWST. Its UV-visible imaging remains uniquely valuable for tracing hot stars, star-formation, and rest-frame ultraviolet structures that are inaccessible to JWST at shorter wavelengths \citep[\eg][]{Windhorst2026}. 

\vfill\break In the area of gravitational lensing, HST has contributed to landmark discoveries, including the first multiply imaged supernova, SN Refsdal \citep{Kelly2015}. HST also observed Earendel, the most distant known star \citep{Welch2022} with a photometric redshift of $z_{\rm phot} \simeq 6.2$ 
\citep[see also][]{Welch2022b} , surpassing the previous record holder, Icarus \citep{Kelly2018}, at \(z \simeq 1.49\) which HST discovered in the same host galaxy as SN Refsdal. 
 
Building on these lensing capabilities, JWST has provided crucial insights into cosmic expansion through the discovery of Supernova (SN) H0pe at \textit{z} = 1.78, a multiply imaged Type Ia SN \citep{Frye2024}. This rare event was discovered when the JWST NIRCam was imaging the galaxy cluster PLCK G165.7+67.0 at $z = 0.348$ \citep{Pascale2022}. The phenomenon of multiply imaged supernovae was first theorized by \citet{Refsdal1964}, who proposed that the multiple light curves produced could be used to infer time delays between images. Combined with an accurate mass distribution model for the lensing cluster, these delays could constrain the Hubble Constant \(H_0\). Following this methodology, \citet{Pascale2024} calculated \(H_0 = 75.7^{+8.1}_{-5.5} \text{ km s}^{-1} \text{ Mpc}^{-1}\), producing a precision measurement that adds to our understanding of cosmic expansion rates. 

This study focuses on the MACS J0416.1-2403 field, hereafter MACS0416, to search for overlooked high-redshift caustic transits. At \(z=0.397\), this lensing cluster is one of the most extensively observed fields and provides a useful opportunity to search for transient sources across multiple epochs. In an arc nicknamed ``Spock," at \(z = 1.0054 \pm 0.0002\) \citep[\eg][]{Rodney2018}, HST discovered two fast transients and an additional transient was detected in an arc straddling a caustic at $z = 0.9397$ \citep{Kaurov2019, Chen2019}. Six additional transients were discovered by the HST ``Flashlights" program \citep{Kelly2022}. More recently, JWST observations of MACS0416 revealed 14 transients in  \citet{Yan2023, Williams2026}. These HST and JWST detections motivate this search which aims to detect additional high-redshift, \(7<z<17\), caustic transits in MACS0416. In total, 23 caustic transits have been reported in MACS0416, with the highest-redshift one being a ``kaiju" star nicknamed Mothra $z=2.091$ from the JWST PEARLS sample \citep{Diego2023, Yan2023}. 

In \S \ref{Data}, we describe the observational data used in this search, and in \S \ref{Methods}, we outline the approach taken to identify possible overlooked high redshift caustic transits in MACS0416. In \S \ref{sky_SB}, we discuss setting upper limits for the cadence of Pop III caustic transits, walk through the process of updating the $2~\mu$m Pop III sky SB in the \citet{Windhorst2018} framework, and our initial steps for a model dependent estimate of the $2~\mu$m Pop III sky SB. With the help of cluster caustics, the JWST is in a prime position to study objects from the Epoch of First Light. We adopt a flat $\Lambda$CDM cosmology with $\text{H}_0=67.66 \text{ km} \text{ s}^{-1} \text{Mpc}^{-1}$, $\Omega_m=0.30966$, and $\Omega_b = 0.04897$. All magnitudes are quoted in the AB system \citep{Oke1983}, and all coordinates are given in the ICRS frame (equinox J2000).

\section{Data}\label{Data}
\indent This search uses data from the Prime Extragalactic Areas for Reionization and Lensing Science (PEARLS; PID 1176 and 2738; PI R. Windhorst) JWST GTO program \citep{Windhorst2023}, which includes a strong time domain component and observations of the lensing cluster MACS0416. The mosaics were produced following the methodology first described by \citet{Koekemoer2011}, including updates for JWST. The data were reduced using version 1.13.4 of the JWST Science Calibration Pipeline with Calibration Reference Data System (CRDS) with context 1230. Custom corrections were applied to mitigate $1/f$ noise and other low-level detector artifacts.

JWST/NIRCam observed this cluster in 3 epochs total for the PEARLS program, allowing for a wide range of time-domain astronomy, making it a particularly attractive field for high-redshift caustic transit detection. MACS0416 was observed on October 7th 2022, December 29th 2022, and February 10th 2023 for a total of 126 days between observations. The Canadian NIRISS Unbiased Cluster Survey (CANUCS; \citet{Willott2022}) covered MACS0416, but was omitted from this study as it occurred only 13 days after the PEARLS Epoch 2 and registration of CANUCS epoch onto PEARLS epoch 2 was problematic.

The 3 PEARLS NIRCam observations, Epoch 1, Epoch 2, and Epoch 3 (ep1, ep2, and ep3 hereafter), were made in the same eight filters. The short wavelength filters are F090W, F115W, F150W, and F200W. The long wavelength filters are F277W, F356W, F410M, and F444W. Although NIRCam's short and long-wavelength detectors have native pixel scale of 0.031$^{\prime\prime}$ pixel\(^{-1}\) and 0.063$^{\prime\prime}$ pixel\(^{-1}\), respectively, all images used here were drizzled to a common pixel scale of 0.031$^{\prime\prime}$ pixel\(^{-1}\). All MACS0416 observational data can be found in Table 2 of \citet{Windhorst2023} and a summary found in Table \ref{tab:obs}

\vspace{-4cm}
\setlength{\tabcolsep}{17.5pt}
\begin{deluxetable}{lcc}[ht!]
\tablecaption{JWST NIRCam observations of MACS\,J0416.1$-$2403 \label{tab:obs}}
\tablehead{
\colhead{Filter} & \colhead{Exp. time (s)} & \colhead{Mean AB depth}
}
\startdata
F090W & 3779 & 28.47 \\
F115W & 3779\tablenotemark{a} & 28.47 \\
F150W & 2920 & 28.49 \\
F200W & 2920 & 28.70 \\
F277W & 2920 & 29.85 \\
F356W & 3779\tablenotemark{a} & 29.90 \\
F410M & 3779\tablenotemark{a} & 29.51 \\
F444W & 3779 & 29.51 \\
\enddata
\tablenotetext{a}{One of the three epochs utilized a shorter exposure time (3349\,s for F115W and F410M; 2920\,s for F356W).}
\tablecomments{Epochs: 1 = 2022-10-07, 2 = 2022-12-29, and 3 = 2023-02-10. Listed depths are the mean 5$\sigma$ depths over the three epochs in AB-mags, measured from the RMS map within circular apertures of radius 0.2\arcsec.}
\end{deluxetable}
\setlength{\tabcolsep}{13pt}

\vspace{-5cm}
NIRCam has two flanking modules, A and B, both of which were used throughout all observations. Module B was chosen as the primary module for transit detection, as it is spatially overlapped in all three epochs. The short wavelength band consists of 4 detectors, and an \texttt{INTRAMODULEBOX} was used to cover the gaps between the detectors. All PEARLS observations of MACS0416 used the ``up-the-ramp'' fitting technique, and the \texttt{MEDIUM8} readout pattern was adopted. 

\begin{figure}[H]
    \begin{minipage}[t]{0.47\textwidth}
        \centering
        \includegraphics[width=\linewidth]{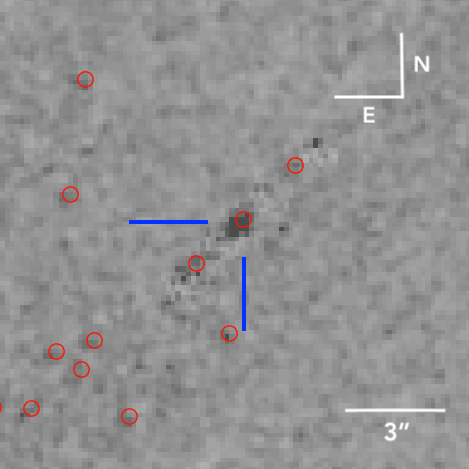}
        \caption{This transient is one of 12 detected in \citet{Yan2023} and serves as validation for this detection technique with red circles being detected sources. The transient was found in the arc at $z=2.091$ of MACS0416. The image uses an inverted color map, where darker regions indicate positive flux differences. In this difference-image, $\mathrm{diff}_{31}$, the transient is brighter in Epoch 3 than in Epoch 1.}
        \label{fig:mothra}
    \end{minipage}
    \begin{minipage}[t]{0.47\textwidth}
        \centering
        \includegraphics[width=\linewidth]{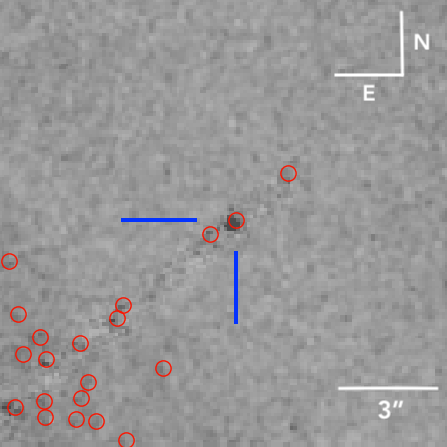}
        \caption{
            This source (shown in $\mathrm{diff}_{31}$), detected in the Spock arc of MACS0416 and reported in \citet{Yan2023}, serves as a second validation of our detection technique. }
        \label{fig:spock}
    \end{minipage}
\end{figure}
\vspace{-0.5cm}

\section{Methods}\label{Methods}

\indent The science images were constructed and difference-images were made by direct subtraction. The F356W filter was used as the primary detection filter as it provides the deepest data available in our set, with a mean \(5\sigma\) depth of 29.90 mag across the three JWST PEARLS observations of MACS0416, see Table \ref{tab:obs}. Other filters were reserved for verifying candidate reliability.

The longer wavelength filters are less prone to the short wavelength artifacts such as wisps, making F356W an ideal starting point for visual inspection. Each difference-image was inspected by eye because unresolved residuals and image processing artifacts produce many spurious detections. Wisps were removed following the procedure in \citet{Windhorst2023}. 

To limit the time required for visual inspection, difference-image combinations were limited to $\rm ep3-ep1$ ($\rm diff_{31}$), $\rm ep2-ep3$ ($\rm diff_{23}$), and $\rm ep2-ep1$ ($\rm diff_{21}$). These image combinations were used to detect sources brightening in Epoch 3, fading in Epoch 2, and brightening in Epoch 2, respectively.

\subsection{Image Preparation and Source Extraction}
\indent Each difference-image combination used a combined weight map, see Figure \ref{fig:weight}, to optimize faint source detection in \texttt{Source Extractor} \citep{Bertin1996}.

\noindent The combined weight map was constructed from the inverse variance maps of the ``parent images'' and inverted to yield the final combined weight map. While this approximation does not fully account for the correlated pixel noise in the drizzled mosaics, it is sufficient for our analysis since the weight map was used primarily to minimize spurious detections during visual inspection. Weight maps store inverse variance estimates (\(1/\sigma^2\)) for each pixel, which \texttt{Source Extractor} converts to position-dependent noise levels. 

From source-masked blank-sky apertures of radius=3.33 pixels in $\text{diff}_{31}$, we derived a $5\sigma$ in-aperture limiting magnitude of 29.04 AB. Using the F356W encircled-energy fraction \textit{EE}=0.753, this corresponds to an aperture-corrected total point-source sensitivity of  28.73 AB mag. There is approximately $\sqrt{2}$ more noise in the difference image than the parent image, ideally, the point source detection limit in the difference image should be $\approx0.4 $mag shallower than the individual F356W epoch depths. Factors like correlated drizzle noise, imperfect source masking, and background residuals push the point-source detection limit in the difference image shallower. This could potentially explain the difference between the ideal $5\sigma$ detection limit and the one found. 

With the flux from the science image and the position-dependent noise levels, a pixel-wise Signal to Noise Ratio (SNR) can be calculated, essentially telling the observer how much confidence one can have in each pixel. In addition, point spread functions (PSFs) were generated with WebbPSF \citep{Perrin2015} and supplied to the extraction pipeline \texttt{Source Extractor}.

\begin{center}
\begin{minipage}{0.47\textwidth}
    \centering
    \includegraphics[width=\linewidth]{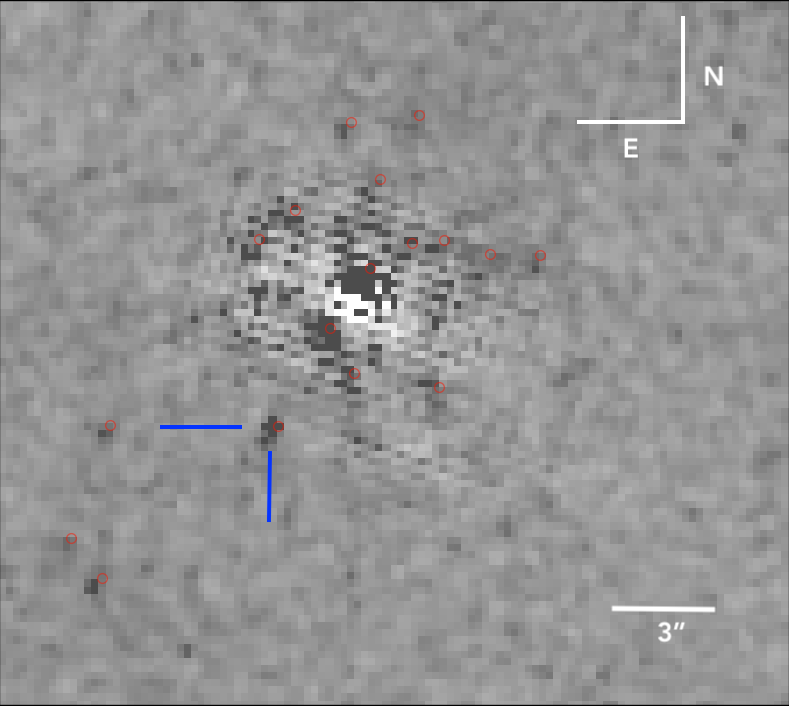}
    \captionsetup{width=\linewidth}
    \captionof{figure}{
        Possible candidate detected in $\mathrm{diff}_{31}$ in an off-center region of MACS0416. The source is located near the edge of the mosaic and is only covered by two exposures, one of which contains a hot pixel at the source's location, confirming that the detection is unreliable, and likely a hot pixel or residual cosmic ray.
    }
    \label{fig:cand}
\end{minipage}
\end{center}

This WebbPSF-generated model accounts for instrument-specific characteristics, including diffraction spikes and segmented mirror gaps, improving faint source detection, particularly in crowded fields where false positives are abundant. During source extraction, we used a single PSF rather than convolving the parent images to a common PSF. The image rotation between the 3 epochs is well sampled, and the PSF in each of the 3 epochs are very similar. Because our pipeline relies on rigorous visual inspection of all candidates, any minor residuals from unmatched PSFs (such as residual diffraction spikes) were easily identified and discarded. Using a single PSF significantly expedited the extraction process without compromising the detection of real transit sources. Position angles for $\rm ep1, ep2, \text{and }  ep3$ are $293^\circ, \ 33^\circ, \text{and } 71^\circ$ respectively. Angles are measured from the detector y-axis projected onto the sky, in degrees east of north.

\indent Before the visual inspection stage our detection method needed to be validated.
To do so, the Mothra transient \citep{Diego2023, Yan2023} and the transient in the Spock Arc \citep{Rodney2018, Caminha2017} needed to be detected by \texttt{Source Extractor}, see Figure \ref{fig:mothra} and \ref{fig:spock}, respectively. Before detecting both transits, \texttt{Source Extractor} parameters needed to be fine-tuned to extract faint to nearly invisible sources. 
At the beginning of the source detection phase, Source Extractor  would routinely detect \(\gtrsim 10^5\) sources in each difference-image. After adjusting the necessary parameters and adding the combined weight images and the PSF, \texttt{Source Extractor}  would detect $\lesssim 10,000$ sources in each difference-image. The decreasing number of extracted sources and the detection of the two ``validation transits'' proved that our \texttt{Source Extractor} parameters were viable for faint source detection.

Two criteria were applied during detection process: sources with a Full Width at Half Maximum (FWHM) $\leq$ 2 pixels and a \texttt{MAG\_AUTO} (Kron-like automated aperture magnitude in AB mag) $\geq$ 70 were excluded. Sources with a FWHM $\leq$ 2 were excluded (\(\approx 10^4 \text{ sources}\)) to reduce spurious detections, since single pixel noise spikes or cosmic rays can mimic compact point sources. Sources with a \texttt{MAG\_AUTO} $\geq 70$ were excluded, as this threshold effectively filters out the default error flags (\texttt{MAG\_AUTO}=99.0) output by \texttt{Source Extractor} for invalid measurements. These cuts reduce the visual inspection workload and remove instrumental artifacts from further analysis, without excluding slightly extended sources because the F356W PSF FWHM is 3.7 pixels meaning real sources should survive the cut. Overall, the exclusion thresholds prioritize real caustic transit candidates while minimizing false positives during visual inspection.

\subsection{Visual Inspection and False-Candidate Detection}
Running \texttt{Source Extractor} on an image produces an output catalog containing information on each extracted source. \texttt{GlueViz} \citep{Robitaille2017} was used for data exploration and for linking the \texttt{Source Extractor} catalog to the difference-image. Visual inspection was performed systematically by following critical curves from our strong lensing model \citep{Bergamini2023a} at the relevant redshift values. This approach was applied across all difference-image combinations, with each \texttt{Source Extractor} catalog linked to its corresponding image.

While examining the F356W diff\(_{31}\), we identified a possible variable candidate at (R.A.\ $14^\mathrm{h}16^\mathrm{m}04\fs022$, Decl.\ $-$24\degr04\arcmin35\farcs37) with MAG\_AUTO = 29.50 ( Figure~\ref{fig:cand}). This candidate lies far from the cluster's core, does not appear to be multiply imaged, and shows a positive flux residual in the difference-image, making it a plausible transient candidate rather than a lensed caustic event. 

\begin{figure}[h!]
    \centering
    \includegraphics[width=1.00\textwidth]{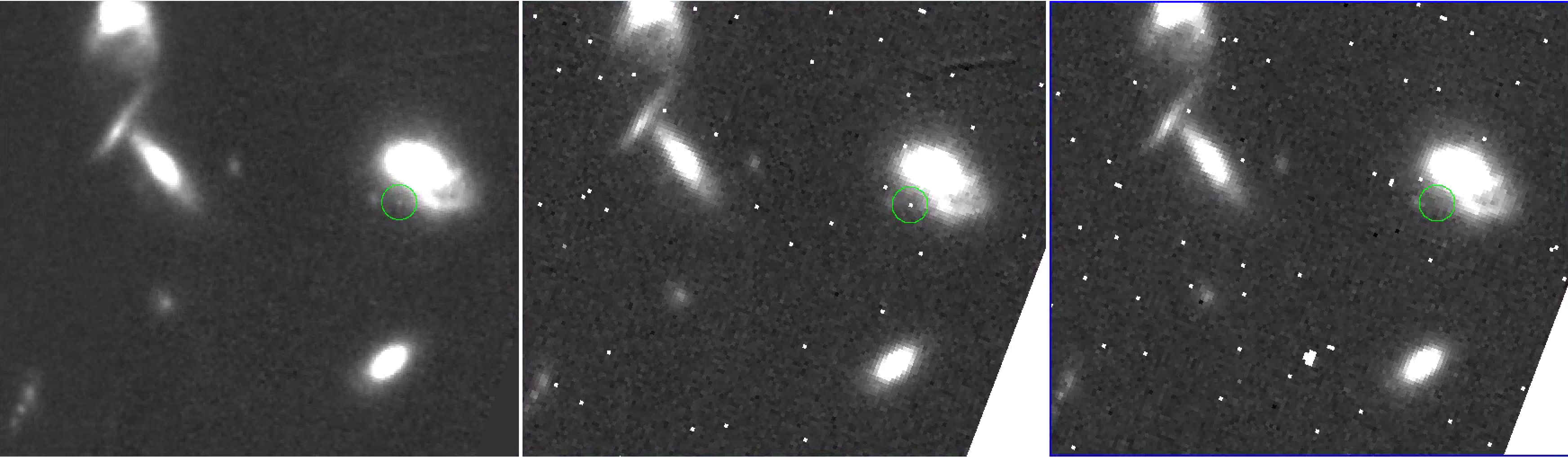}
    \includegraphics[width=1.00\textwidth]{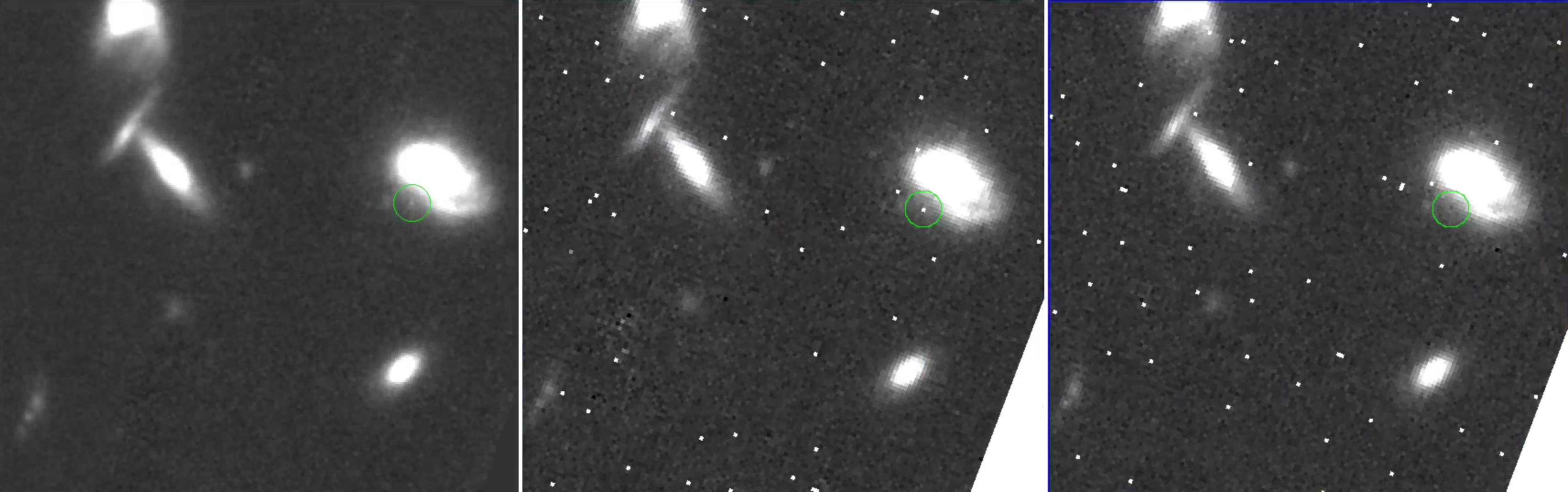}
    \captionsetup{width=1.00\linewidth}
    \parbox{\textwidth}{\caption{Footprints of the NIRCam F277W (top row) and F356W (bottom row) filters at the location of the false candidate. From left to right, the three columns show, in order, the final mosaic, the first exposure, and the second exposure in each filter. The ``false candidate" (green circle) is present only in the final mosaic and is located at a pixel that is hot in one of the individual exposures of each filter, demonstrating that it is a hot pixel rather than a natural source. In the outer parts of the mosaic, where only two exposures cover the region, rejection of the hot pixel leaves a single contribution exposure corresponding to a lower weight, making such artifacts more common in the flanking regions of the field.}
    \label{fig:277W}}
\end{figure} 

The candidate was initially considered a supernova, but was later shown to be an artifact. The source drops out in all other filters except for F277W, which, like F356W only has two exposures covering the source region. One of the exposures contains a hot pixel exactly at the candidate's location (Figure \ref{fig:277W}), indicating that the detection is caused by a variable hot, high dark-current pixel rather than a real source. 

With the visual inspection of all difference-image combinations yielding no high redshift (z $\gtrsim 7$) sources and the false candidate being refuted, our analysis constrains the expected rate of caustic transits in observations of lensing clusters. These constraints inform not only the frequency of such rare events, but also place empirical limits on the underlying Pop III star surface density and illustrates how future JWST time-domain observations can be optimized to improve the likelihood of detection. 
  
\section{Constraints on the 2~\micron\  Sky-SB and Caustic Transits of Pop III stars}
\label{sky_SB}

In this section, we summarize the existing constraints on the 2 $\mu$m Pop III star sky-SB, as well as the expected cadence of Pop III star caustic transits.

\subsection{Constraints on the 2~\micron\ Pop III star sky-SB}

\citet{Windhorst2018} and \citet{Cooray2012} give consistent estimates for the \(2\mu\)m sky-SB, the latter reporting \(\gtrsim31.91\) mag \(\text{arcsec}^{-2}\) and the former 
\(\geq31.4\pm0.6\) mag \(\text{arcsec}^{-2}\). Here, we will only consider the quantity from \citet{Windhorst2018}. This limit, derived from Panchromatic Extragalactic Background Light (EBL) measurements and near-infrared power spectra, represents the maximum amount of unresolved light that may be contributed by Pop III stars and their remnants without contradicting observational evidence. We refer to \citet{Windhorst2018} for more information on the derivation of this limit. 

We focus on two key parameters in deriving the limit to the Pop III star surface density from the lack of caustic transit detections behind MACS 0416: caustic length and the detection limit. In \citet{Windhorst2018}, a caustic length of 100\arcsec \ and a conservative JWST NIRCam detection limit of 28.5 mag were adopted. The $5\sigma$ point source sensitivity limit in our F356W difference-image is 28.73 AB mag, a modest 0.23 mag deeper than the assumed detection limit above. We adopt the \citet{Bergamini2023a} model as our fiducial lensing model for the caustic transit search, for which the mean tangential caustic length found for redshifts 7, 12, and 17 is $293.7\pm33$\arcsec. We omit the contribution from radial caustics, since near the clusters center microlenses heavily suppress the maximum magnification making transit detection challenging. To estimate the model dependence of the mean tangential caustic length, we also considered the updated version of our chosen parametric model from \citet{Rihtarsic2025}, which includes new JWST constraints and gives a mean caustic length of 311.7\arcsec, as well as an independent free form model from \citet{Diego2024}. This gives a mean tangential caustic length of 245.7\arcsec \ over the same redshift range. Taking half of the full spread among these redshift averaged caustic lengths, we adopt a model dependent uncertainty of $\pm$33\arcsec,which corresponds to a magnitude error of 0.12 mag which was found from:

\begin{equation}
    \Delta\rm m = 2.5log_{10}\left(\frac{293.7+33}{293.7}\right).
    \label{eq:causterr}
\end{equation}

The total caustic length in the source plane is one important factor controlling how many individual stars experience high magnification, and thus helps set the expected number of caustic crossing events for a given surface density of stars. Assuming the same stellar luminosity distribution as in \citet{Windhorst2018}, the expected caustic crossing event rate scales with both the caustic length and projected stellar surface density in the regions probed by the caustics.

Despite the longer caustics and our slightly deeper imaging, no caustic transits were detected. This may imply that the background population is intrinsically fainter or significantly less dense than previously assumed. Another possibility is that the background stars could possess exceptionally large radii, which would lower their peak magnification. Rather than increasing the actual photometric depth of the image, the longer caustic length increases the effective source-plane area probed for highly magnified stars. We quantify this increased probability as a scaling factor applied to the surface brightness limit, yielding a magnitude difference of \(2.5\log_{10}(\frac{293.7}{100})\approx 1.17\) mag. When combined with the 0.23 mag improvement in our photometric detection limit, the overall constraint on the background population improves by $1.17+0.23=1.4 $ mag. Applying this directly to the reference case above, constrains the 100 $M_\odot$ 2 $\mu$m Pop III sky SB to \(\geq32.8\pm0.61\) mag \(\text{arcsec}^{-2}\), including the uncertainty propagated from the caustic length.

\subsection{Interpreting the Pop III Non-Detection}

\citet{Windhorst2018} provided a reference case discussing the sky-SB if there were 1000 Pop III stars each of 100\Msol\ per $\text{arcsec}^2$, then their integrated 2.0 $\mu\text{m}$ sky-SB would be \(\gtrapprox33\) mag \(\text{arcsec}^{-2}\). Our updated 2.0 $\mu\text{m}$ sky-SB of \(\geq32.8\pm0.6\) mag \(\text{arcsec}^{-2}\) is only $\sim0.2$ mag from testing the reference case. To improve the SNR of the difference-image, PSF-matched detection could be implemented. Stacking difference-images could also be used, as long as the time between repeat observations was short compared to the source's variability timescale. 

Assumed in our Pop III 2.0 $\mu\text{m}$ sky-SB estimate is that the clusters transverse velocity is $1000 \text{ km } \text{s}^{-1}$, which was the adopted value used in deriving the EBL based 2.0 $\mu\text{m}$ sky-SB \citep{Windhorst2018} and is also predicted by \citet{Kelly2018} for MACS0416 specifically. Both papers also consider the possibility of transverse velocities reaching $\leq2000 \text{ km s}^{-1} $. For PLCK G165.7 + 67.0, cluster members have transverse velocities of $2000{}^{+3800}_{-2000} \text{ km s}^{-1}$ and $2200\pm1500 \text{ km s}^{-1}$ according to Nonrelativistic Hydrodynamic Flow Models, and $3000\pm1200 \text{ km s}^{-1}$ and $1800\pm650 \text{ km s}^{-1}$ according to Mach Cone Models \citep{Wilde2026}.

If we were to assume that MACS0416 has a transverse velocity of $2000 \text{ km s}^{-1}$ and take the assumption from \citet{Windhorst2018} that caustic transit rate scales linearly with transverse velocity, adopting $2000\text{ km s}^{-1}$ instead of $1000\text{ km s}^{-1}$ would shift our 2.0 $\mu\text{m}$ sky-SB limit fainter by 0.75 mag to \(\geq33.55\pm0.61\) mag \(\text{arcsec}^{-2}\) making it comparable to the reference case. Using $10^{-0.4(m_1-m_2)}$ we find the sky-SB flux ratio between the reference case of $33 \text{ mag arcsec}^{-2}$ and the updated \(\geq33.55\pm0.61\) mag \(\text{arcsec}^{-2}\) which is equal to $.60$ mag. In the \citet{Windhorst2018} framework, this would correspond to a projected Pop III surface density of $\lesssim 600 \text{ stars arcsec}^{-2}$, rather than $1000 \text{ stars arcsec}^{-2}$. If the projected Pop III surface density of $\lesssim 600 \text{ stars arcsec}^{-2}$ is assumed to be distributed uniformly in comoving density over $7<z<17$, then for our flat $\Lambda$CDM cosmology this corresponds to a light-cone averaged comoving number density of $1.4\times10^2 \text{ cMpc}^{-3}$.

We use Bayesian inference for a Poisson process to update the caustic transit rate per cluster, \(\lambda\), given our non-detection. Adopting a conjugate Gamma prior, \(\lambda \sim {\rm Gamma}(\alpha,\beta)\), with \(\alpha_{\rm prior}=1\) and \(\beta_{\rm prior}=1/0.32=3.125\) yr so that the prior mean is \(0.32\ \mathrm{yr}^{-1}\), the posterior for \(k=0\) events in an effective observing time of \(t\simeq0.34\) yr is:

\vspace{-3ex}
\begin{equation}
\lambda \mid {\rm data} \sim {\rm Gamma}(1,3.465).
\label{eq:eq4}
\end{equation}

This gives a posterior mean of \(0.29\ \mathrm{cluster}^{-1}\,\mathrm{yr}^{-1}\) caustic transits, corresponding to roughly one expected Pop III caustic transit per 3.5 cluster-years of monitoring. More importantly for the non-detection case, the 95\% upper credible limit, defined by \(P(\lambda < \lambda_{95}\mid {\rm data})=0.95\), is \(\lambda_{95}=0.87\ \mathrm{cluster}^{-1}\,\mathrm{yr}^{-1}\). Our rate constraint is tied to the $100 M_{\odot}$ \citet{Windhorst2018} reference case, and therefore applies to the detectable caustic transits at an assumed JWST single-epoch depth of approximately AB$\simeq28.5 $ mag.

As an additional test, we analyzed the residual pixel noise statistics in source free regions bounded by critical curves to search for unresolved faint source emissions. After masking resolved sources, the remaining pixels values were binned by source redshift and magnification. The populated bins contained between 139 and 4,335 unmasked pixels, with median residual values typically ranging from $\approx 3.5\times10^{-2}$ to $5.6\times10^{-2}$ nJy $\text{pixel}^{-1}$. The fraction of pixels above the median by $3\sigma$ was $<1.5\%$, with only a few bins reaching values near 2\%. We find no strong systematic excess, although some high magnification bins exhibit weak positive tail deviations that may warrant further investigation. The limited number of unmasked pixels in the highest magnification bins increases the uncertainty in the deviation measurements. The positive tail deviation could be explained by factors like residual light from incompletely masked sources, Intracluster light, correlated drizzle noise.

\subsection{Toward Stellar-Model-Dependent Caustic-Transit Constraints}
  
The conversion from the expected number of detectable caustic transits to surface brightness, or to the source-plane surface number density of stars, is model-dependent. Mechanisms that regulate these relations include the cosmic star-formation history of Pop III stars, the Pop III stellar initial mass function, the evolution of Pop III stars during their lifetimes (including the effects of binary evolution and stellar rotation), the nebular emission from re-processed ionizing radiation from Pop III stars, and accretion onto the black holes that the Pop III stars leave behind. Since the interplay between these different effects is highly non-trivial, we will here adopt the \citet{Windhorst2018} scaling between Pop III surface number density, surface brightness, and transit detection rate -- which assumes single, non-rotating Pop III stars and no nebular emission.

As an initial step towards a stellar model dependent calculation, not considering microlenses, we evaluated the detectability of individual stellar evolutionary track points using the Muspelheim \citep{Zackrisson2024} synthetic photometry for the \citet{Windhorst2018} $10{\rm \, M_{\odot}}$ -- $100{\rm \, M_{\odot}}$ Pop III stellar models. Muspelheim also includes a zero metallicity $10{\rm \, M_{\odot}}$ -- $1000{\rm \, M_{\odot}}$ Pop III model \citep{Yoon2012}, a $9{\rm \, M_{\odot}}$ -- $120{\rm \, M_{\odot}}$ model \citep{Murphy2021}, and a $100{\rm \, M_{\odot}}$ -- $1000{\rm \, M_{\odot}}$ model \citep{Volpato2023}. For every initial mass, source redshift, and evolutionary time step, we use the unlensed JWST/NIRCam F356W magnitudes and calculated the minimum magnification needed to reach the empirical $5\sigma$ F356W difference-image detection limit of $m_{\rm lim}=28.73$ AB mag: 
\begin{equation}
    \mu_{\rm min}
    =
    10^{0.4\left(m_{\rm F356W}-m_{\rm lim}\right)}.
    \label{eq:mumin}
\end{equation}
The F356W band is used as the model selection band because the null result presented in this work is based on the F356W difference-image search. 

We convert $\mu_{\rm min}$ into an idealized source plane detection zone around a fold caustic, a locally smooth segment of the caustic. Adopting the local point source relation:
\begin{equation}
    \mu(d_\perp)=\frac{B_0}{\sqrt{d_\perp}},
    \label{eq:foldcaustic}
\end{equation}
where $d_\perp$ is the perpendicular source plane distance from the caustic in arcsec and $B_0$ is the local fold caustic strength, the maximum distance at which a point source attains the threshold magnification is:  
\begin{equation}
    d_{\rm max}
    =
    \left(\frac{B_0}{\mu_{\rm min}}\right)^2.
    \label{eq:dmax}
\end{equation}

\noindent We evaluate this quantity for values of $B_0=10$ and 20 \citep{Miralda-Escude1991, Diego2017, Windhorst2018}, which are nominal values for Hubble Frontier Fields like MACS0416.

We then perform an approximate finite-source peak-reachability screen. For a stellar radius $R_\star$, the source angular radius is $\theta_\star=R_\star/D_A(z)$, converted to arcsec. In the local caustic approximation, the maximum finite-source magnification scales approximately as:
\begin{equation}
    \mu_{\rm max}
    =
    \frac{B_0}{\sqrt{\theta_\star}}, 
    \label{eq:mumax}
\end{equation}
where $\theta_\star$ is in arcsec. Track points satisfying $\mu_{\rm min}\leq\mu_{\rm max}$ are classified as potentially peak reachable under the adopted smooth caustic assumptions. This screening step captures the competing effects of stellar evolution: luminous evolved stars can require lower $\mu_{\rm min}$, whereas their larger radii can lower $\mu_{\rm max}$.  We also define the peak reachable time fraction as the fraction of the evolutionary time represented by the available track during which a star can, in principle, reach the magnification required for F356W detection. This smooth caustic calculation neglects foreground micro lensing and uses representative values of $B_0$, rather than a spatially varying caustic strength. It should therefore be interpreted as a first-order peak reachability screen rather than a complete event-selection model. 

The preliminary calculation demonstrates a strong mass dependence, see Fig \ref{fig:finitesource} and Table \ref{tab:finitesource_z7}. Among the \citet{Windhorst2018} track points, the most favorable $100\,M_\odot$ models at $z=7$ require $\mu_{\rm min} \approx9.2\times10^3$, while the most favorable $10\,M_\odot$ models require $\mu_{\rm min} \approx1.9\times10^6$. Under the finite-source screen, all stellar evolution track model points (typically sampled at every 10$^5$ years) at initial masses of $15$--$100\,M_\odot$ are reachable during the peak of the caustic transit for both $B_0=10$ and 20. In contrast, the $10\,M_\odot$ track is sensitive to the assumed caustic strength. For $B_0=10$, the time fraction reachable at the peak decreases from 0.60 at $z=7$ to 0.065 at $z=17$. For $B_0=20$, all sampled $10\,M_\odot$ stellar evolution track model points are reachable at the caustic transit peak. 

A complete stellar-model-dependent $2 \, \mu\rm m $ sky-SB constraint will require convolving these individual-star calculations with a Pop III IMF, a redshift-dependent Pop III SFH, the time represented by each evolutionary track point, and the recovery probability associated with the three epoch F356W difference-image \citep[see \eg][]{Berkheimer2026}. Such a calculation will also require observer-frame $2 \, \mu\rm m $ flux densities for the model tracks. We defer this full forward modeling analysis to future work; the W18-scaled sky-SB limit remains the primary quantitative constraint of the present paper.

\section{Conclusion} \label{Conclusion}

\indent Using three JWST/NIRCam epochs of MACS0416, we searched for high-redshift caustic transits candidates over the redshift internal $7\leq z \leq17$. Previous analyses of this field reported 12 caustic transits \citep{Yan2023} and two additional caustic transits \citep{Williams2026}. However, our search found no additional candidates consistent with Pop III caustic transits in the targeted redshift range. Interpreted within the \citet{Windhorst2018} framework, this null result implies a 95\% upper credible limit on the detectable Pop III caustic transit rate of:
\begin{equation}
    \lambda_{95}
    =
    0.87\ {\rm cluster}^{-1}\,{\rm yr}^{-1},
\end{equation}
compared with the W18 reference estimate of $0.32\ {\rm cluster}^{-1}\,{\rm yr}^{-1}$.

Using the W18 scaling between the Pop III source density, caustic transit rate, and integrated near-infrared emission, we constrain the observer-frame $2~\mu{\rm m}$ Pop III sky SB to be fainter than \(32.8\pm0.61\) mag \(\text{arcsec}^{-2}\), based on the slightly deeper $5\sigma$ detection limit of our difference-images and the longer caustic length relative to the assumption adopted by \citet{Windhorst2018} in their derivation. Under the additional assumption that the caustic transit rate scales linearly with transverse velocity, adopting a fiducial cluster transverse velocity of \(2000\ \mathrm{km \ s}^{-1}\) instead of \(1000\ \mathrm{km \ s}^{-1}\) shifts this limit to \(\geq 33.6 \pm 0.6 \text{ mag arcsec}^{-2}\), making it directly comparable to the \citet{Windhorst2018} reference case, allowing us to update the projected Pop III surface density to $\lesssim 600 \text{ stars arcsec}^{-2}$. 

We also preformed an initial stellar-model-dependent assessment of the individual source selection underlying this interpretation. Using Muspelheim synthetic photometry \citep{Zackrisson2024} for the W18 Pop III stellar tracks, we calculated the F356W magnification needed for individual evolutionary track points to reach the empirical difference-image detection threshold and applied an approximate finite-source peak-reachability screen. This calculation demonstrates a strong mass dependence: $100\,M_\odot$ track points at $z=7$ require magnifications of order $10^4$, whereas favorable $10\,M_\odot$ points require magnifications of order $10^6$. The finite-source reachability of the $10\,M_\odot$ models is sensitive to the assumed caustic strength, while the sampled $15$--$100\,M_\odot$ tracks remain reachable under both of the adopted smooth caustic scenarios. This preliminary calculation does not yet replace the W18-scaled sky SB constraint, because a complete model-dependent estimate requires Pop III IMF and SFH weighting, and finite source light curves, and a cadence dependent recovery calculation. 

These results support the rarity of first-light objects and demonstrate the value of time-domain observations of massive lensing cluster for constraining stellar populations that remain far below nominal JWST detection limits. Future monitoring with additional epochs, shorter temporal separations, and comparable depth across multiple massive cluster fields will improve sensitivity to short lived caustic transit events, constrain the time-dependent selection function, and enable more stringent stellar-model-dependent limits on the integrated Pop III $2~\mu{\rm m}$ sky SB.

\section{Acknowledgments}
 This work is based on observations made with the NASA/ESA/CSA James Webb Space Telescope. The data were obtained from the Mikulski Archive for Space Telescopes (MAST) at the Space Telescope Science Institute, which is operated by the Association of Universities for Research in Astronomy, Inc., under NASA contract NAS 5-03127 for JWST. These observations are associated with JWST programs 1176 and 2738. Supported in part through the Arizona NASA Space Grant Consortium, Cooperative Agreement 80NSSC20M0041 or NSSC25M7084. EZ acknowledges project grant 2022-03804 from the Swedish Research Council.

We also acknowledge the indigenous peoples of Arizona, including the Akimel O'odham (Pima) and Pee Posh (Maricopa) Indian Communities, whose care and keeping of the land has enabled us to be at ASU's Tempe campus in the Salt River Valley, where much of our work was conducted.

\bibliography{m}
\bibliographystyle{aasjournal}

\onecolumngrid
\nolinenumbers

\clearpage
\appendix

\indent PSFs were used to mitigate erroneous source detection during source extraction. PSF convolution was outside the scope of this study. It did not appear to have a negative impact on source detection, as the goal of introducing the PSF into Source Extractor  was to reduce the number of detections, which were brought down from \(\approx 100,000\) to \(\approx 9,000\) for each difference-image combination.  

\begin{figure}[!htbp]
    \includegraphics[width=1\linewidth]{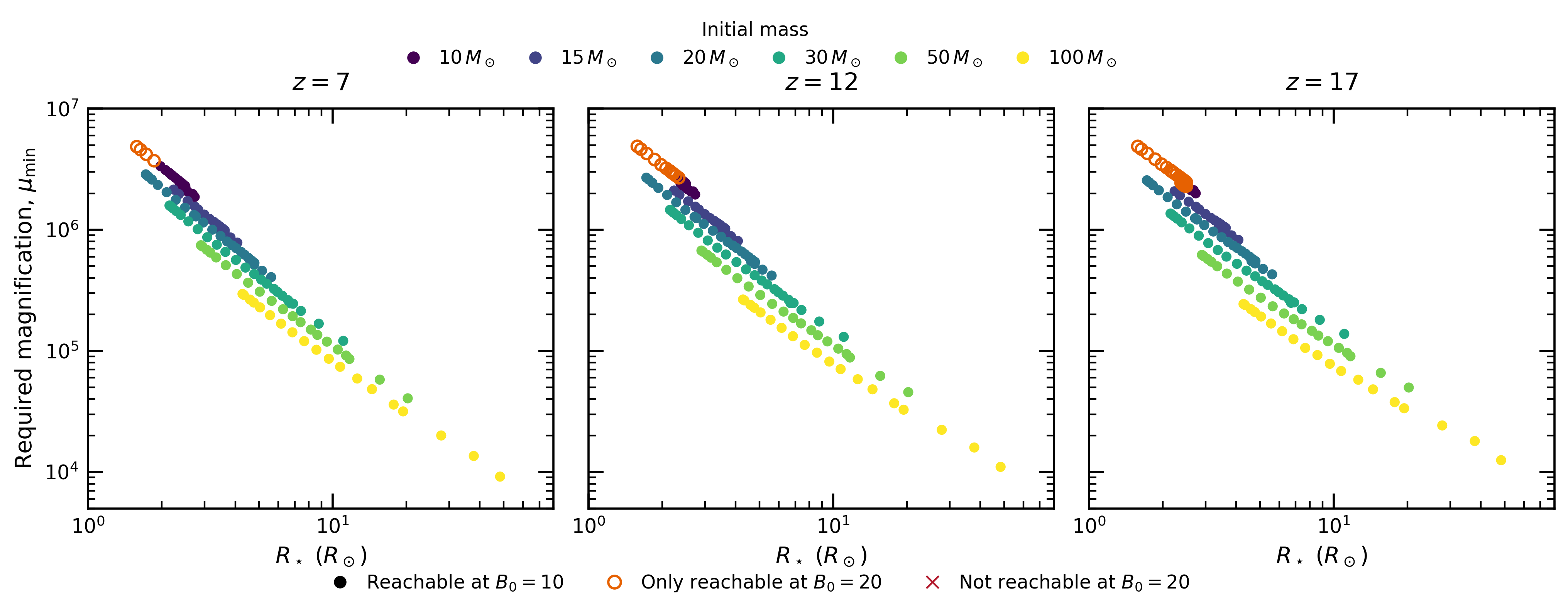}
    \caption{Minimum F356W magnification required for detection, $\mu_{\rm min}$, as a function of stellar radius for W18-derived Pop III stellar evolutionary track points evaluated using Muspelheim synthetic photometry. The three panels show source redshifts $z=7$, 12, and 17. Colors indicated initial stellar mass. Filled symbols satisfy the approximate finite-source peak-reachability criterion for a fold caustic strength of $B_0=10$. Open orange circles fail this criterion for $B_0=10$ but satisfy it for $B_0=20$; red crosses fail the criterion for both values. The figure demonstrates the strong dependence of individual star F356W caustic-transit detectability on initial stellar mass, evolutionary state, source redshift, and the adopted local caustic strength.}
    \label{fig:finitesource}
\end{figure}

\noindent\makebox[\textwidth][c]{%
\begin{minipage}[t]{0.78\textwidth}

\refstepcounter{table}
\label{tab:finitesource_z7}

{\textbf{Table \thetable.} Representative F356W caustic transit
detectability metrics at $z=7$ for the W18-derived Pop~III track grid.\par}

\vspace{0.5ex}

\centering
\small
\setlength{\tabcolsep}{6pt}

\begin{tabular*}{\linewidth}{@{\extracolsep{\fill}}rrrrrr@{}}
\hline
$M_{\rm init}$ &
Track span &
Best $m_{\rm F356W}$ &
Lowest $\mu_{\rm min}$ &
Best $d_{\rm max,B10}$ &
$f_{\rm peak,B10}$ \\

($M_\odot$) &
(Myr) &
(AB mag) &
&
(arcsec) &
\\
\hline
10  & 8.208 & 44.408 & $1.87\times10^{6}$ & $2.87\times10^{-11}$ & 0.603 \\
15  & 4.026 & 43.464 & $7.83\times10^{5}$ & $1.63\times10^{-10}$ & 1.000 \\
20  & 7.892 & 42.752 & $4.06\times10^{5}$ & $6.06\times10^{-10}$ & 1.000 \\
30  & 5.878 & 41.436 & $1.21\times10^{5}$ & $6.84\times10^{-9}$  & 1.000 \\
50  & 4.219 & 40.250 & $4.06\times10^{4}$ & $6.08\times10^{-8}$  & 1.000 \\
100 & 3.026 & 38.634 & $9.15\times10^{3}$ & $1.19\times10^{-6}$  & 1.000 \\
\hline
\end{tabular*}

\vspace{0.75ex}

\raggedright
\small
\textit{Note.} The track span is the age interval represented by the
available input grid, not necessarily the complete stellar lifetime. Best
values refer to the tabulated evolutionary point with the smallest required
F356W magnification. The peak-reachable time fraction is the fraction of the
tabulated evolution satisfying the approximate finite-source
peak-reachability criterion for $B_0=10$.

\end{minipage}%
}

\begin{figure}[H]
    \centering
\includegraphics[width=.40\linewidth]{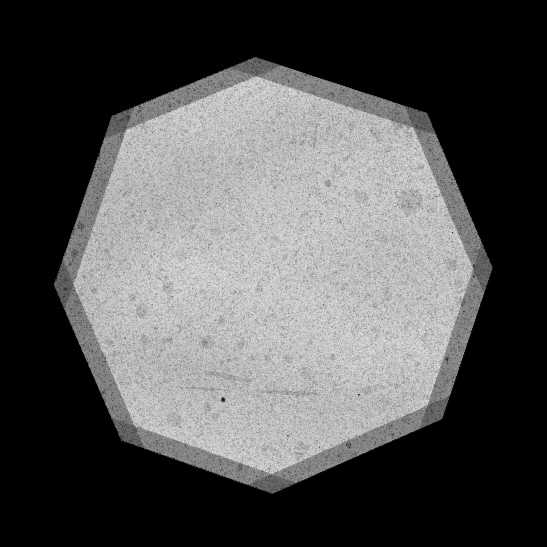}
    \caption{
        The combined weight map from the overlapping region of ep3 and ep1. Because of the additive nature of weight values, pixels in spatially overlapped regions are given a higher confidence value since $\sigma_{\text{combined}} = \frac{\sigma}{\sqrt{N}}$, where $N$ is the number of exposures.
    }
    \label{fig:weight}
\end{figure}

\vspace{-1cm}
\begin{figure}[H]
    \centering
    \includegraphics[width=.40\linewidth]{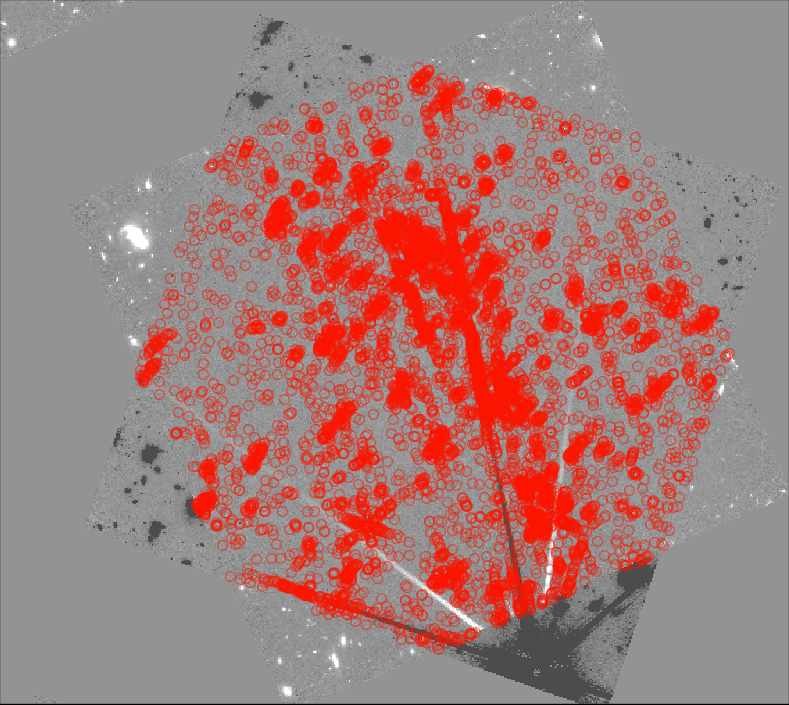}
    \caption{
        difference-image (diff$_{31}$) with an inverted color map where each circle indicates an extracted source \(\approx\) 9000. Each source close to a critical curve of interest was visually inspected to determine its validity, as visual inspection is one of the more reliable methods for detecting positive flux differences, especially at such high redshifts. Using \texttt{glueviz}, the \texttt{Source Extractor } catalogs' were matched to difference-images WCS which greatly improved time spent during the visual inspection stage.
    }
    \label{fig:diff_image}
\end{figure}

\noindent\makebox[\textwidth][c]{%
\begin{minipage}[t]{0.72\textwidth}

\refstepcounter{table}
\label{tab:SourceExtractor_f356w_simple}

{\textbf{Table \thetable.} \texttt{Source Extractor} configuration for the
F356W transit search\par}

\vspace{0.5ex}

\centering
\small
\setlength{\tabcolsep}{5pt}

\begin{tabular*}{\linewidth}{@{\extracolsep{\fill}}llll@{}}
\hline
Parameter & Value & Parameter & Value \\
\hline
\texttt{DETECT\_MINAREA}  & 4 &
\texttt{THRESH\_TYPE}     & RELATIVE \\

\texttt{DETECT\_THRESH}   & 1.90 &
\texttt{ANALYSIS\_THRESH} & 1.90 \\

\texttt{FILTER\_NAME}     & \texttt{gauss\_2.0\_5x5.conv} &
\texttt{DEBLEND\_NTHRESH} & 64 \\

\texttt{DEBLEND\_MINCONT} & 0.005 &
\texttt{MAG\_ZEROPOINT}   & 28.0865 \\

\texttt{BACK\_SIZE}       & 64 &
\texttt{BACK\_FILTERSIZE} & 3 \\

\texttt{WEIGHT\_TYPE}     & MAP\_WEIGHT &
                            & \\
\hline
\end{tabular*}

\vspace{0.6ex}

\raggedright
\small
The listed parameters define the \texttt{Source Extractor} configuration
used for the F356W difference-image candidate search. Candidate validation
relied on subsequent visual inspection and multi-filter checks.

\end{minipage}%
}

\clearpage
\end{document}